\documentclass{article} 

\usepackage{cogsci}

\cogscifinalcopy 

\usepackage{hyperref}
\usepackage{pslatex}
\usepackage{apacite}
\usepackage{graphicx}
\usepackage{natbib}
\usepackage{amsmath}
\usepackage{float} 

\usepackage{url}
\usepackage{booktabs}

\usepackage{siunitx}
\usepackage[table]{xcolor}
\usepackage{array}
\usepackage{multirow}
\usepackage{caption}
\usepackage{geometry}
\usepackage{tabularx}
\usepackage{longtable}

\title{Validating Generative Agent-Based Models of Social Norm Enforcement: From Replication to Novel Predictions}

\author{Logan Cross$^1$, Nick Haber$^3$, Daniel L.K. Yamins$^{1,2}$ \\$^1$Department of Computer Science, $^2$Department of Psychology, $^3$Graduate School of Education, Stanford University}

\begin{document}

\maketitle

\begin{abstract}
As large language models (LLMs) advance, there is growing interest in using them to simulate human social behavior through generative agent-based modeling (GABM). However, validating these models remains a key challenge. We present a systematic two-stage validation approach using social dilemma paradigms from psychological literature, first identifying the cognitive components necessary for LLM agents to reproduce known human behaviors in mixed-motive settings from two landmark papers, then using the validated architecture to simulate novel conditions. Our model comparison of different cognitive architectures shows that both persona-based individual differences and theory of mind capabilities are essential for replicating third-party punishment (TPP) as a costly signal of trustworthiness. For the second study on public goods games, this architecture is able to replicate an increase in cooperation from the spread of reputational information through gossip. However, an additional strategic component is necessary to replicate the additional boost in cooperation rates in the condition that allows both ostracism and gossip. We then test novel predictions for each paper with our validated generative agents. We find that TPP rates significantly drop in settings where punishment is anonymous, yet a substantial amount of TPP persists, suggesting that both reputational and intrinsic moral motivations play a role in this behavior. For the second paper, we introduce a novel intervention and see that open discussion periods before rounds of the public goods game further increase contributions, allowing groups to develop social norms for cooperation. This work provides a framework for validating generative agent models while demonstrating their potential to generate novel and testable insights into human social behavior.

\end{abstract}

\section{Introduction}

Advances in LLMs have created new possibilities for studying human social behavior through simulation. While agent-based modeling has long been used in the social sciences, the emergence of LLM-based agents enables a qualitatively new approach to simulate humans: generative agent-based modeling (GABM)  \citep{vezhnevets2023generative, anthis2025llm}. 

Generative agents are LLM-powered entities that combine multiple cognitive components—such as memory, personality, and reasoning modules—through structured prompt sequences to simulate human decision-making processes \citep{park2023generative,park2024generative}. These agent architectures function as computational models of human cognition, similar to the "box-and-arrow" diagrams used in cognitive psychology \citep{neisser2014cognitive}. For GABMs to be useful tools for social science, we need rigorous frameworks to validate whether these agents actually capture relevant aspects of human behavior. By presenting agents with the same experimental stimuli used in human studies, we can directly test whether their stimulus-to-behavior mappings match those observed in human participants. Thus, in this paper we introduce a two-stage approach that prototypes GABMs to replicate and extend human studies. \footnote{
Code: \url{https://github.com/locross93/Multi_Agent}
}

\textbf{Model Validation:} First, we conduct an iterative model comparison to identify the minimal agent architecture (set of cognitive components) sufficient to reproduce known behavioral effects in complex social settings. Each “model” in our framework represents a different LLM agent architecture—specifically, different combinations of prompts which we call cognitive components. These components are added to a base agent who takes in the responses to these prompts as context for making a decision, for example the answer to “what kind of person am I?” provides additional context to guide the agents decisions about which actions to take \citep{march2008logic,vezhnevets2023generative}. Our validation of each model works as follows: We run the same experimental protocol used in the original human studies by representing the experimental stimuli in natural language and compare our agents’ behavioral patterns to the published human data using identical statistical tests. An architecture is considered “validated” when agents show the same directional effects as humans with statistical significance. For theoretical parsimony in line with suggested best practices for GABM \citep{vezhnevets2023generative}, we begin with the simplest possible agent architecture and test new cognitive components only when simpler models fail to capture human behavior patterns. In addition, by varying the agent architectures and comparing their output to human data, we can test specific hypotheses about the cognitive mechanisms underlying social behavior.

\textbf{Novel Prediction Generation:} Second, we leverage validated agent architectures to explore counterfactual scenarios that would be slow or costly to implement in laboratory settings. By maintaining the validated agent architectures while varying environmental conditions, we can generate precise, quantitative predictions about human behavior in novel situations. These predictions can then be tested empirically, providing a rigorous test of the model's generalization capabilities while advancing our understanding of human social behavior.

Here, we present this validation approach with two case studies to examine the components necessary for GABMs to replicate key findings from experimental social dilemmas. Mixed-motive social dilemmas, particularly the trust game and public goods games, represent the gold standard for experimental research on social dynamics across psychology, economics, and social sciences because they uniquely capture the fundamental tension between individual self-interest and collective welfare that defines human social life \citep{van2013psychology}. Unlike approaches that study cooperation or competition in isolation, these paradigms model the complex mixed incentives that characterize most important social challenges — from maintaining personal relationships \citep{crocker2017social} to climate change cooperation \citep{milinski2008collective} to following cultural norms \citep{koster2022spurious,leibo2024theory}. Greater theoretical frameworks of social dilemmas help us to understand why people cooperate (or not) and also practical ways to maintain and promote cooperation in groups and organizations through coordination and social norm enforcement. This research has revolutionized our understanding through methodologically rigorous frameworks that bridge laboratory precision with real-world relevance \citep{fehr2000cooperation, henrich2006costly}. First, we investigate third-party punishment (TPP) in the trust game from \citet{jordan2016third}, which demonstrated that third-party punishers are perceived as—and actually are—more trustworthy than non-punishers. Through careful comparisons of different agent architectures built in Concordia, we show that both persona prompting and theory of mind reflection are necessary components for reproducing these effects. Second, we examine \citet{feinberg2014gossip}'s work on gossip and ostracism in promoting cooperation in a public goods game, finding that while our initial validated architecture from the TPP study provides a good foundation, an additional strategic component proves necessary to replicate a majority of the behavioral effects in this more complex social environment.

We then apply these validated model architectures to explore two distinct types of novel questions not investigated in the original papers. For the TPP study, we use the validated model for theoretical disambiguation, examining how punishment behavior differs between public and private settings. Our results quantify the relative contributions of reputational vs. intrinsic motivations, helping resolve the ongoing theoretical debate about why people engage in costly third-party punishment. For the public goods study, we use the validated model to test a novel intervention—adding pre-round discussion periods—and find that this further improved cooperation rates, as groups could self-organize and establish shared norms for contribution. These findings demonstrate how GABM can help social scientists both evaluate conflicting cognitive theories and investigate potential interventions to improve collective welfare.

\section{Related Work}

Recent advances in large language models have enabled promising simulations of human behavior \citep{anthis2025llm}. Generative agent-based modeling (GABM) has emerged as a useful approach that leverages LLMs' capabilities in natural language interaction, commonsense reasoning, and episodic memory \citep{vezhnevets2023generative}. \citet{park2023generative} demonstrated that generative agents can produce believable simulations of human behavior and social networks, while follow-up work by \citet{park2024generative} expanded these simulations to include agents that could predict the experimental behavior of 1,000 humans they interviewed. This influential research focuses on modeling specific individual differences. In contrast, rather than focusing on individual prediction, our work produces complex social phenomena from the dynamic interaction of individual agents and validates at the population effect level. Our work establishes a rigorous framework for validating whether generative agents can accurately reproduce specific behavioral effects from controlled human experiments and the cognitive mechanisms necessary to do so. The Concordia framework \citep{vezhnevets2023generative} was specifically designed to facilitate such social simulations, providing researchers with tools to ground agent interactions in physical, social, and digital spaces. Beyond simulations, other researchers have explored persona-based prompts to enhance diversity in agent behavior \citep{li2023camel, chen2024persona}, which is critical to representing individual differences at the population level in decision making.

The two experimental paradigms used in this study represent core problems in experimental economics. The Trust Game is a classic paradigm first introduced by \citet{berg1995trust} to measure trust and trustworthiness through sequential monetary exchanges. Third-party punishment, as examined by \citet{jordan2016third}, extends this paradigm to include unaffected observers who can pay costs to sanction norm violations, raising questions about both reputational and intrinsic motivations for such behavior \citep{jordan2020signaling}. The Public Goods Game used in our second study is a fundamental paradigm for studying cooperation in groups \citep{ledyard1994public}. When participants can contribute to a shared resource that benefits everyone, the dominant strategy for self-interested individuals is to contribute nothing while benefiting from others' contributions. This creates a social dilemma where individual rationality leads to collective irrationality.

\begin{figure*}
\centering
\includegraphics[width=1.0\textwidth]{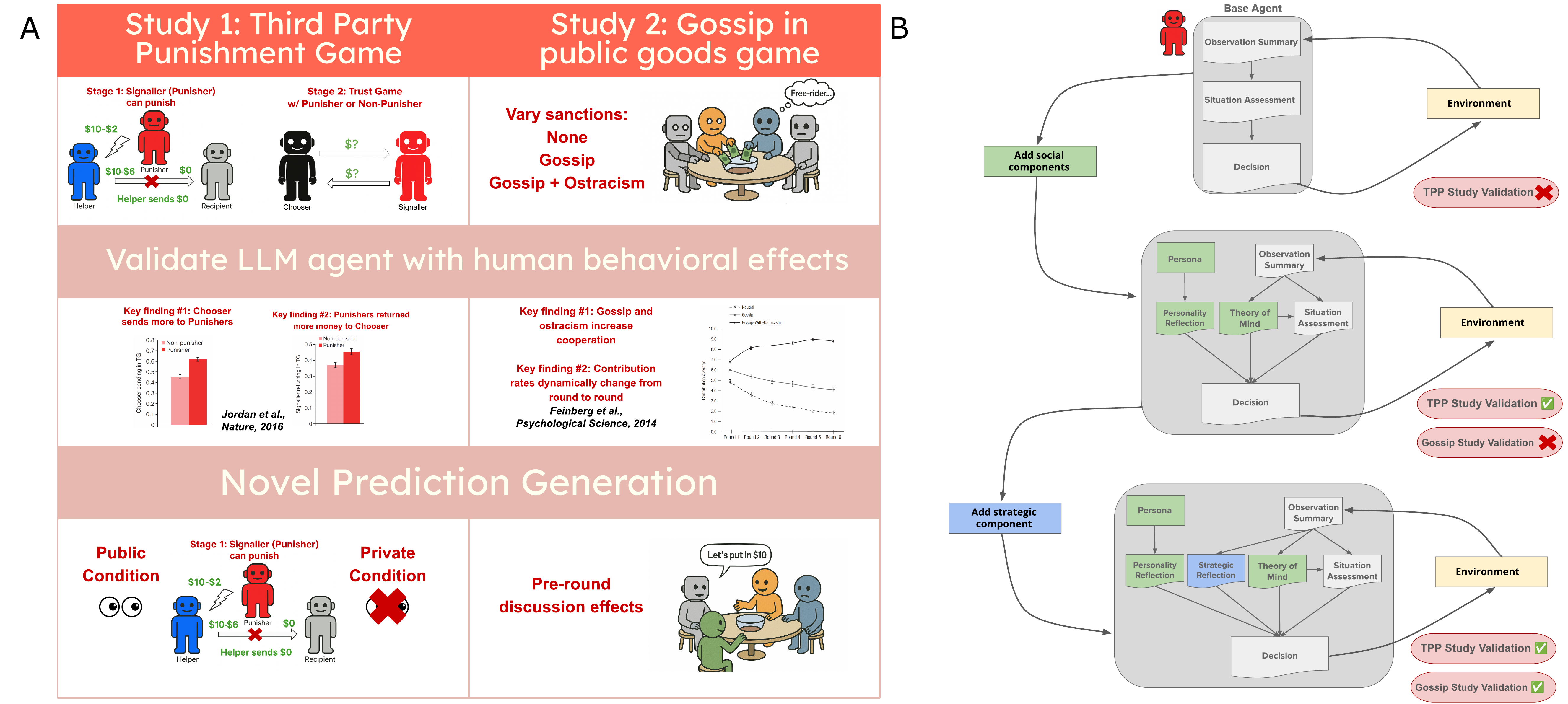}
\caption{A. Two-stage approach of model validation and novel prediction generation. B. Iterative refinement of model architectures.} 
\label{fig2_new}
\end{figure*}

\section{Methods}

We implemented our experimental paradigms using Concordia \citep{vezhnevets2023generative}, a framework that enables LLM-based generative agents to interact through natural language while maintaining grounded state variables for monetary amounts and game decisions. Concordia's flexible component-based architecture allowed us to create modular cognitive capabilities as prompt response chains, each with its own state and dependencies on other components. The framework dynamically constructs the prompt that conditions the final LLM call by concatenating responses from cognitive components according to the agent's architectural configuration.

Formally, each cognitive component $i$ produces text output from an LLM call based on its dependencies and the agent's memory of all observations:
\begin{equation}
z_i^{t+1} \sim p(\cdot | f^i(\{z_j^t : j \in \text{dep}(i)\}, m_t))
\end{equation}
where $f^i$ is a formatting function that concatenates responses from components that component $i$ depends on, given the agent architecture configuration (Figure 1B), and $m_t$ represents the agent's accumulated memory up to time $t$. 

In the action step, the agent samples its decision $a_t$ from an LLM call given a contextual prompt that concatenates the outputs of all cognitive components with directed edges into the decision component:

\begin{equation}
a_t \sim p(\cdot | f^a(\{z_j^t : j \in \text{dep}(\text{action})\}))
\end{equation}
where $\{z_j^t : j \in \text{dep}(\text{action})\}$ represents the current states of cognitive components with directed edges into the decision component, and $f^a$ is a formatting function that concatenates these component outputs into the context prompt conditioning the LLM's action selection.

This modular design enables systematic hypothesis testing about which cognitive processes are necessary and sufficient for reproducing human social behavior. By systematically varying which components are included in \text{dep}(\text{action}) while maintaining identical experimental protocols, we can isolate the causal contribution of different cognitive mechanisms to observed behavioral patterns.

\subsubsection{Base Agent Architecture}
All agent architectures include the fundamental components necessary for participating in experimental protocols:
\textbf{Observation Summary}: Processes information about current and past game states to feed to a prompt to summarize all the observations. This includes a memory subcomponent that maintains record of all the past interactions an agent observed. The output is also prefixed by a list of all observations, altogether forming $m_t$.

\textbf{Situation Assessment Component}: Evaluates the current decision context using the prompt "What is the current situation that [agent name] faces in the [Game Type]?" This component depends on the Observation Summary output and provides situational awareness.

\textbf{Decision}: Prompts the LLM to generate a decision following additional chain of thought. In the each architecture, the output of the other components with directed edges into the decision component are concatenated in the prompt.

\subsubsection{Additional Cognitive Components}

We systematically varied the inclusion of additional cognitive components to identify minimal architectures for replicating human behavior:

\textbf{Persona Component}: Contributes stable individual differences through demographic background and personality traits. Uses the prompt "Based on this background and your past actions, how would you describe your approach to trust and fairness in economic decisions vs maximizing your own payoff?" This component provides consistent behavioral tendencies across decisions.
\textbf{Theory of Mind Component}: Enables explicit reasoning about others' mental states and likely responses. Prompts agents to consider "what kind of person their interaction partner is and how they might respond" based on observed behavior patterns.
\textbf{Strategic Reflection Component}: Evaluates long-term payoff maximization through the prompt "What's the best strategy for maximizing your long-term earnings in this game?" This component explicitly considers strategic implications of decisions.
\textbf{Emotion Reflection Component}: Captures affective responses using "How are you feeling emotionally about the current situation?" This component models emotional reactions that may influence decision-making.

\subsubsection{Agent Architecture Configurations}
We tested multiple agent architectures by varying $\text{dep}(\text{action})$
configurations:

\begin{itemize}
    \item Base Architecture: Decision component depends only on Observation Summary and Situation Assessment
    \item Social Architecture: Adds Persona and Theory of Mind components to dependencies
    \item Social + Strategic Architecture (Study 2): Additionally includes Strategic Reflection component
    \item Social + Emotion Architecture (Study 2): Additionally includes Emotion Reflection component
\end{itemize}

This systematic approach allows precise identification of which cognitive mechanisms are necessary for reproducing specific aspects of human social behavior across different experimental paradigms. All agents use GPT-4o as the underlying LLM.

\subsection{Study 1: Third-party Punishment as a costly signal of trustworthiness}

We based our first generative agent simulation on the third-party punishment game developed by \citep{jordan2016third} which consists of two sequential stages designed to test whether costly punishment serves as a signal of trustworthiness. Here, both stages consist of the Trust game. The Trust game is a classic behavioral economics paradigm where one player (the Helper) can choose to send some portion of their endowment to another player (the Recipient), with the sent amount being tripled by the experimenter. Then, the Recipient decides how much of this tripled amount to return to the Helper. This creates a social dilemma that measures trust and trustworthiness: while both players can benefit from cooperation, the Helper may keep the endowment if they do not trust the Recipient, and the Recipient faces a temptation to defect by keeping the tripled amount.

In the first stage, one LLM agent acts as a third party, the Punisher/Signaller, watching two people play the Trust Game. The Trust Game consists of a Helper with an initial endowment of \$10 that could potentially be shared with another player: the Recipient. In Stage 1, the Helper is programmed to give \$0 in order to evaluate whether the Signaller punishes selfish behavior. The Signaller observes the Helper's decision and can pay a personal cost (\$2) to punish the Helper (\$6 reduction of Helper's winnings). The key feature of this stage is that punishment is costly to the Punisher and provides no immediate direct benefit, as they are an unaffected third party.

In the second stage, a new participant (the Chooser) plays a trust game with either a punisher or non-punisher from Stage 1, knowing their previous punishment decision in the public condition (see below). The trust game proceeds as follows: 1. Chooser receives an endowment as the Helper and decides how much to send to the Signaller (Punisher in Stage 1). 2. Any amount sent is tripled by the experimenter. 3. Signaller, acting as the Recipient now, decides how much of the tripled amount to return to the Chooser.

\textbf{Public and Private Conditions}. To disentangle reputational from intrinsic motivations in third-party punishment, we varied punishment decision observability across two conditions. In the public condition (replicating Jordan et al., 2016), the Signaller's punishment decision was explicitly communicated to the Stage 2 Chooser, and Signallers were informed their decision would be public. In the private condition, the Signaller's punishment decision remained unknown to the Chooser, and Signallers were told their decision would remain private. This manipulation removes potential reputational benefits while preserving all other aspects of the decision context.

\subsection{Study 2: Gossip and Ostracism Promote Cooperation in Groups}
We based our second generative agent simulation on the gossip and ostracism paradigm developed by \citet{feinberg2014gossip} which examines how the spread of reputational information through gossip facilitates cooperation and limits defection in groups. The study employed a public goods exercise where participants faced a social dilemma between maximizing personal gain and contributing to group welfare.

\subsubsection{Public Goods Exercise}
The foundation of this study is the public goods exercise. In each round: 1. Participants in groups of 4 each receive an allotment of 10 points (worth 2.5¢ each in the study and \$1 for our agents), 2. Each participant decides how many points to contribute to a group fund versus keep for themselves, 3. The total points contributed to the group fund are doubled and redistributed equally to all group members. This creates a social dilemma where individuals benefit most by free-riding on others' contributions. After each round, participants learned how much each group member had contributed and earned, were assigned to a group with different partners for the next round.

\subsubsection{Experimental Conditions}
The study employed a repeated measures design where all participants played three distinct conditions, each for six rounds: \textbf{Basic} In the basic game, participants played the standard public goods exercise without modifications. \textbf{Gossip} In the gossip condition, after learning the results of each round, participants could send an anonymous gossip note about one of their current group members to that person's future interaction partners. This allowed reputational information to flow between groups despite no direct interaction between them. \textbf{Gossip With Ostracism} This condition added an ostracism mechanism. At the beginning of each round, after receiving any gossip notes, participants could anonymously vote to exclude one participant from playing in the upcoming round. If someone received two or more exclusion votes, they were ostracized and did not participate in that round, earning nothing. The remaining three participants played the public goods exercise with a reduced multiplier (1.5 instead of 2) to maintain proportionality of potential earnings.

\begin{figure*}
\begin{center}
\includegraphics[width=1.0\textwidth]{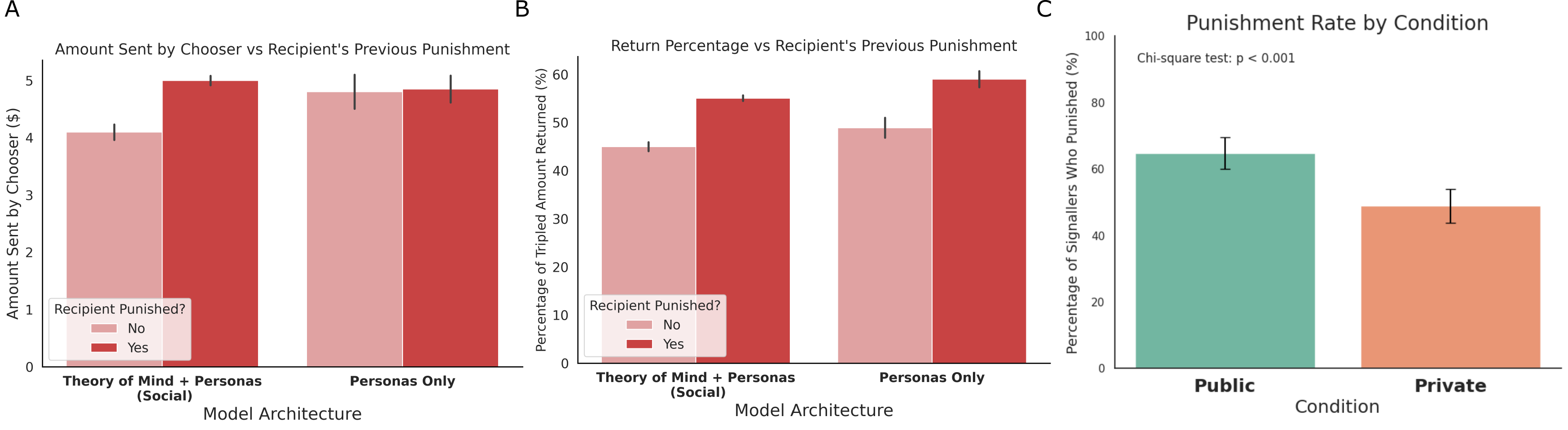}
\end{center}
\caption{TPP Results: (A) Average amount sent by choosers to signallers in Stage 2. (B) Percentage returned by recipients in Stage 2. (C) Punishment rates in public versus private conditions.}
\label{fig3}
\end{figure*}

\section{Results}

\subsection{Study 1: Third-party punishment as a costly signal of trustworthiness}

We first examined what are the minimal components needed in a generative agent model to reproduce key findings from \citet{jordan2016third}'s human study of third-party punishment shown in \autoref{fig2_new}: \textbf{Key Finding 1:} Choosers demonstrated greater trust in punishers by sending them more money in Stage 2. \textbf{Key Finding 2:} Punishers proved to be actually more trustworthy by returning a higher percentage of funds as Recipients in Stage 2.


By systematically varying agent components and comparing their outputs to human data, we can identify which cognitive mechanisms are necessary for reproducing these social behaviors.


\subsubsection{Model Validation}



\textbf{Key Finding 1: Punishment Perceived as a Signal of Trust}. In the original human study, participants consistently demonstrated greater trust in individuals who engaged in costly third-party punishment. Initially, our analyses with the base agent and GPT4o base model showed that persona prompting is necessary to produce these crucial individual differences in punishment rate \autoref{fig2}. Without distinct personas, agents showed uniform punishment rates of 100\%, failing to capture the natural variation seen in human studies, even with a relatively high LLM sampling temperature of 1.0. Thus, adding personas was the first crucial component necessary to model human behavior for this experimental paradigm. 

Without explicit partner modeling, agents failed to discriminate between punishers and non-punishers in their trust decisions. T-tests compare mean amounts sent between groups, where t-values indicate effect size and p-values show statistical significance (t = -0.566, p = 0.572, N=100), sending similar amounts to both groups. Analysis of agent reasoning revealed that without explicit prompting to consider their partner's past behavior, agents defaulted to uniform trust levels regardless of punishment history.  This suggests that explicit partner modeling is necessary for translating punishment signals into trust decisions, mirroring theories about human social cognition.

Following our iterative process, we added a theory of mind component to our agent architecture. With this addition, choosers sent significantly more money (22\% more) to Signallers who had previously engaged in punishment compared to those who did not (t = 6.08, p $<$ 0.001, N = 400) (\autoref{fig3}a, \autoref{table:tpp_stats}). This pattern closely matches the direction and significance found in human studies, though our regression coefficient (9.00) is somewhat smaller than the large effect typically observed in human participants (16.46).

\textbf{Key Finding 2: Punishment is a Signal of Actual Trustworthiness} The second key finding from the human study was that punishers were actually more trustworthy when playing as the Recipient in the trust game.  This relationship demonstrates that punishment isn't merely cheap talk but reflects genuine prosocial tendencies. The Social architecture (ToM + Persona) successfully reproduced this pattern, defined as agents showing the same directional behavioral effect as humans (more trust toward punishers) with statistical significance (p $<$ 0.05): Signallers who had previously punished selfish behavior returned significantly more money to Choosers compared to non-punishers (t = 9.83, p $<$ 0.001, N = 400) (\autoref{fig3}b, \autoref{table:tpp_stats}). This behavioral pattern provided validation that our agents weren't simply learning to signal trustworthiness superficially, but were exhibiting internally consistent patterns of prosocial behavior across different contexts. This pattern persists in the Theory of Mind ablation (t = 6.95, p $<$ 0.001, N=100), as punishers acted less selfishly as Recipients similarly to the Social architecture with ToM (\autoref{fig3}b, \autoref{table:tpp_stats}). This suggests that individual differences implemented through personas are sufficient to drive behavioral consistency through the LLM's modeling of stable prosocial traits. 


\subsubsection{Novel Predictions: Public vs. Private Punishment}


Having validated our Social agent architecture against human behavioral patterns, we used the model to generate and test novel predictions about punishment behavior under public vs. private conditions. Our simulations revealed a significant difference in punishment rates between conditions \autoref{fig3}c. In the public condition, 64.8\% of agents chose to punish the selfish behavior of the Helper in Stage 1, while this rate dropped to 48.9\% in the private condition where punishment decisions were not observable ($\chi^2$ = 18.68, p $<$ 0.001, N = 400). This 16 percentage point difference helps quantify the relative contributions of reputational versus intrinsic motivations for punishment. The 16\% drop in punishment rates when decisions are private represents the portion of punishment motivated by reputational concerns. This suggests that the desire to signal trustworthiness to potential future interaction partners drives a significant portion, but not the majority, of punishment behavior. The substantial baseline rate of punishment in the private condition, where there are no reputational benefits, reveals a strong intrinsic motivation to enforce fairness norms in this task. This suggests that a sizable portion of punishment decisions are driven by genuine prosocial preferences rather than strategic concerns. These quantitative predictions about punishment rates in public versus private conditions offer clear hypotheses for future work with human participants.



\begin{figure*}[t!]
\begin{center}
\includegraphics[width=0.8\textwidth]{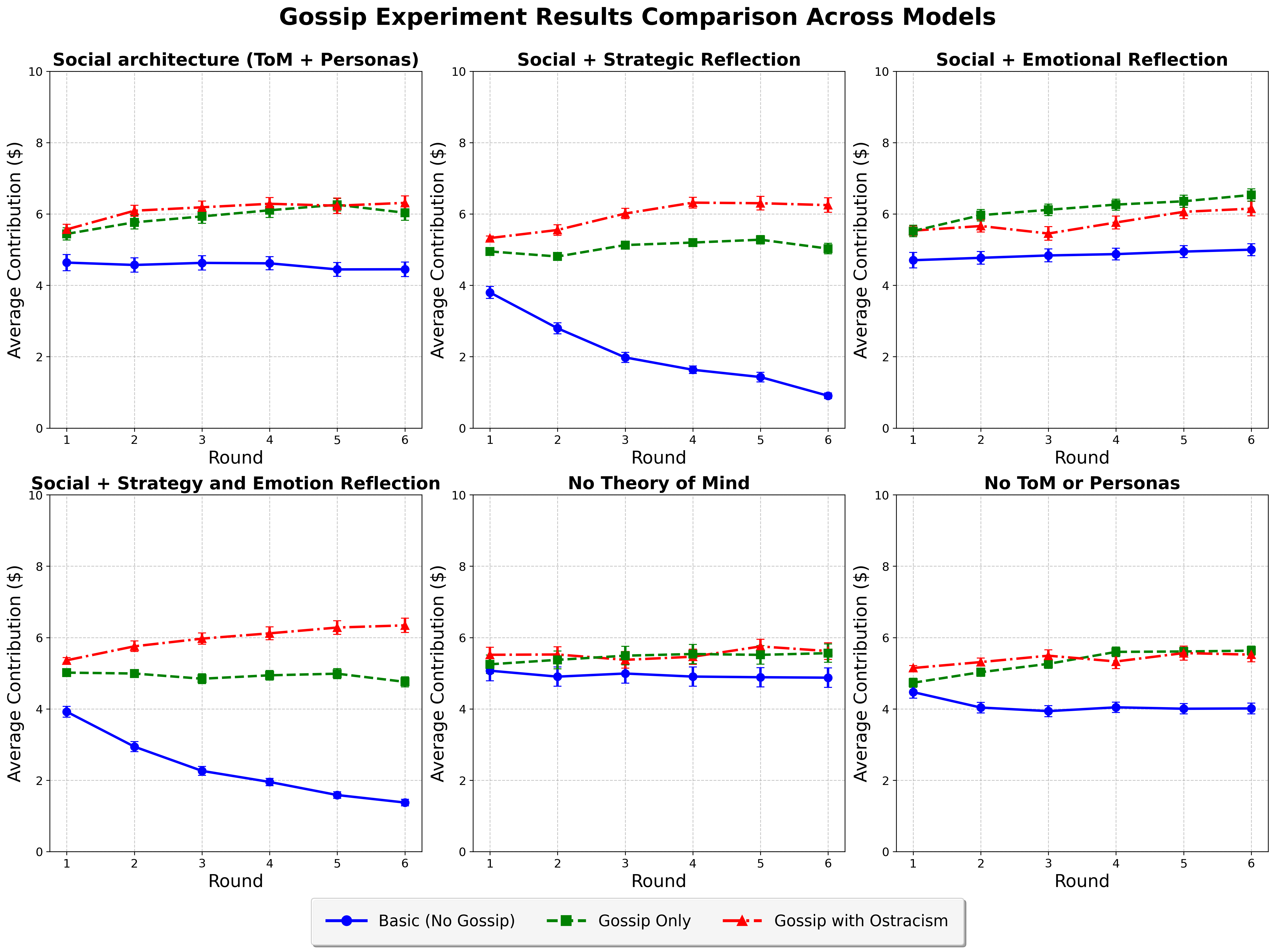}
\end{center}
\caption{Gossip experiment results. Each line includes 5 experiments of 24 agents each.}
\label{fig4}
\end{figure*}

\subsection{Study 2: Gossip and Ostracism Promote Cooperation in Groups}

Building on previous experiments, we investigated the minimal components needed to replicate human social dynamics in \citet{feinberg2014gossip}. In this study, participants in groups of 4 played multiple rounds of a public goods game across three conditions. In the basic condition, participants played a standard public goods game. Two conditions allowed gossip, where participants could write notes about players to their future partners. In the gossip with ostracism condition, players could use this information to vote to ostracize non-cooperative members for a round. Researchers identified key behavioral patterns to replicate: 1. Higher contributions in the gossip + ostracism condition vs basic. 2. Higher contributions in the gossip condition vs basic. 3. Higher contributions in the gossip + ostracism condition vs gossip. 4. Increasing trend in contributions for gossip + ostracism by round. 5. Decreasing trend in contributions for gossip condition by round. 6. Decreasing trend in contributions for basic condition by round. The study demonstrates that gossip and ostracism mechanisms significantly increase group cooperation.

\subsubsection{Model Validation}


We first tested the Social architecture (LLM agent with persona and ToM components) from our previous experiments. This model replicated 3/6 of the experiment effects previously highlighted (\autoref{fig4}, \autoref{table:gossip_stats}, \autoref{table:gossip_trends}). Contributions are significantly higher in both gossip and gossip+ostracism conditions compared to basic ($F(1, 119) = 77.56$, $\eta^2 = 0.39$ for gossip vs. basic; $F(1, 119) = 86.88$, $\eta^2 = .42$ for gossip+ostracism vs. basic). These results show the Social architecture captures reputational transfer effects that function as sanctioning mechanisms increasing overall contributions.


However, the Social architecture fails to replicate other crucial effects from the original study. It doesn't produce the temporal dynamics of declining contributions in the basic and gossip conditions. This effect, common in public goods games \citep{ledyard1994public,fischbacher2001people}, occurs as incentives lead to more free-riding behavior over time. This creates a downward spiral in the basic condition, and to a lesser degree in the gossip-only condition. Even though mutual cooperation yields higher payouts for everyone, free-riding becomes the dominant strategy without effective sanctioning mechanisms like ostracism.

Relatedly, the Social architecture fails to replicate the significant difference between the gossip and gossip + ostracism conditions. In humans, contributions were highest in the gossip and ostracism condition as it creates a social norm in which cooperation is advantageous and norm-breaking is costly. The explicit sanction of ostracism additionally causes a phase shift in temporal dynamics, making cooperation rational and altering the game's equilibrium. Our model shows increasing contributions over time in both the gossip+ostracism condition and the gossip condition, leading to no significant difference between the groups (F(1, 119) = 1.55, $\eta^2 = .01$, p $> 0.05$).


We had two cognitively-grounded hypotheses for improving our model: a strategic reflection component to explicitly consider payoff maximization, and an emotional component to capture risk aversion and affective responses to cooperation/defection.

The strategic component significantly improved model performance (\autoref{fig4}, \autoref{table:gossip_stats}). When added to the social architecture, it enabled clear differentiation between all three conditions, with highly significant pairwise comparisons between conditions. Contributions were highest for gossip-with-ostracism, and significantly higher than the gossip condition ($t(119) = -7.30$, $p < 0.001$), matching human data and showing that ostracism provides benefits beyond reputation alone. The Social + Strategic model also demonstrated appropriate temporal trends—decreasing contributions in the basic condition (slope: $-0.5407$, $p = 0.0017$), and increasing contributions in gossip-with-ostracism (slope: $0.205$, $p = 0.014$)—closely matching human patterns (\autoref{table:gossip_trends}). However, contributions in the gossip condition remained flat rather than decreasing (slope: $0.053$, $p = 0.225$).

In contrast, the emotional reflection component did not improve replication. All conditions actually showed increasing linear trends, and the emotional component model showed higher contributions in the gossip condition compared to gossip-with-ostracism ($t(119) = 2.83$, $p = 0.005$), the opposite of what was observed in the human study. Combining both components yielded results similar to the strategic-only model, suggesting strategic reasoning is the critical cognitive component.

Ablation studies confirmed the importance of our base architecture components, with models lacking theory of mind or personas failing to replicate key behavioral patterns and condition differences. Importantly, our new architecture with strategic reflection maintained successful replication of the key findings from Study 1 (\autoref{table:tpp_stats}), providing additional validation that these cognitive architectures capture essential social dynamics across different experimental paradigms.

\subsubsection{Novel Predictions: Pre-round Discussion Condition}

We next used the Social + Strategic Reflection model to explore a novel intervention: adding structured discussion periods before each round of the public goods game. This condition builds on the gossip-with-ostracism mechanism but adds a collective deliberation phase, allowing agents to coordinate their behavior and establish shared norms explicitly rather than only through indirect gossip channels.

\begin{figure}
\begin{center}
\includegraphics[width=1.0\columnwidth]{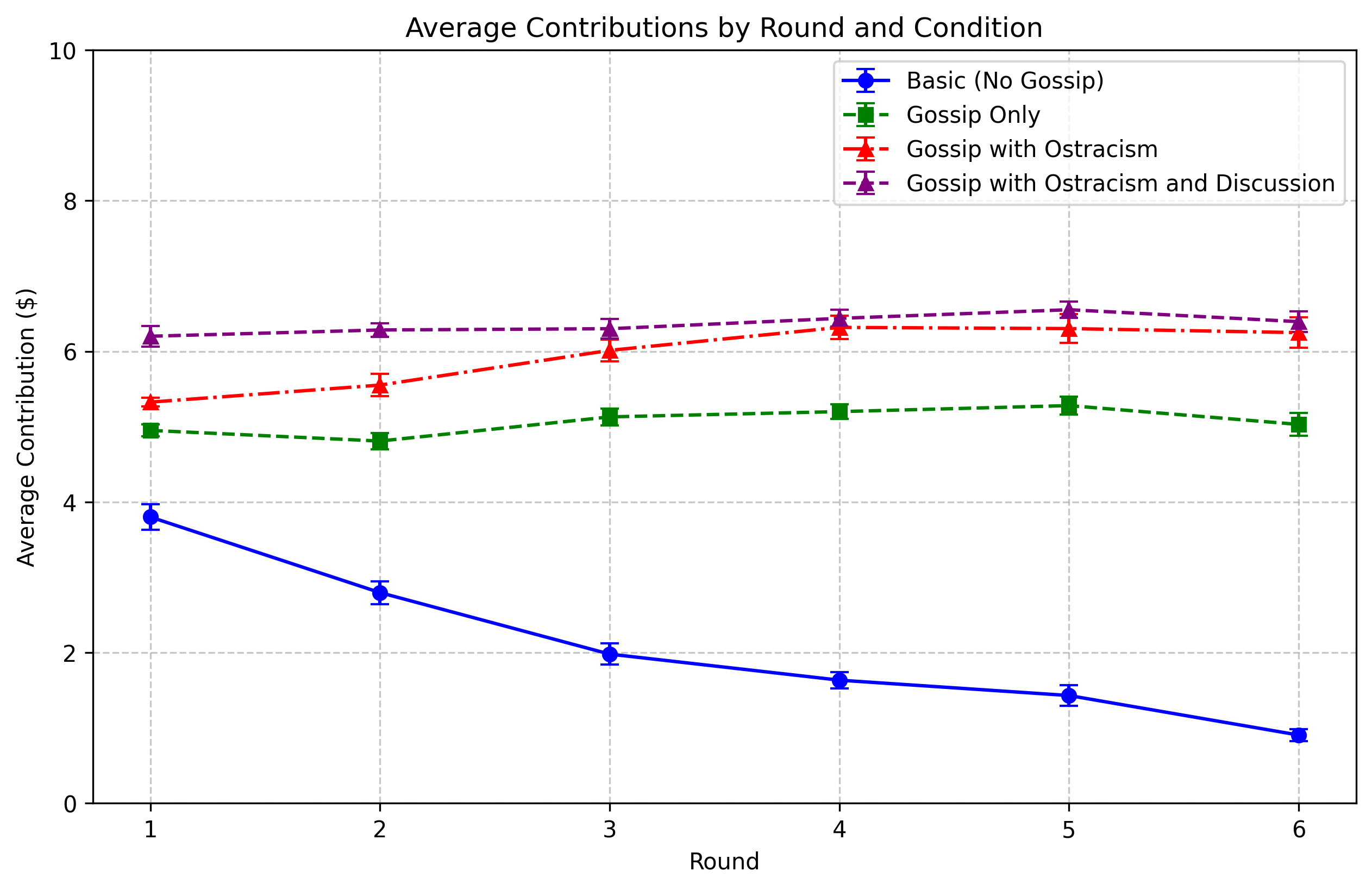}
\end{center}
\caption{Additional condition in Study 2, examining whether an open-discussion period before round would increase contributions. Y-axis shows average contribution by agent} 
\label{discussion}
\end{figure}

\begin{figure}[H]
\begin{center}
\includegraphics[width=0.7\linewidth]{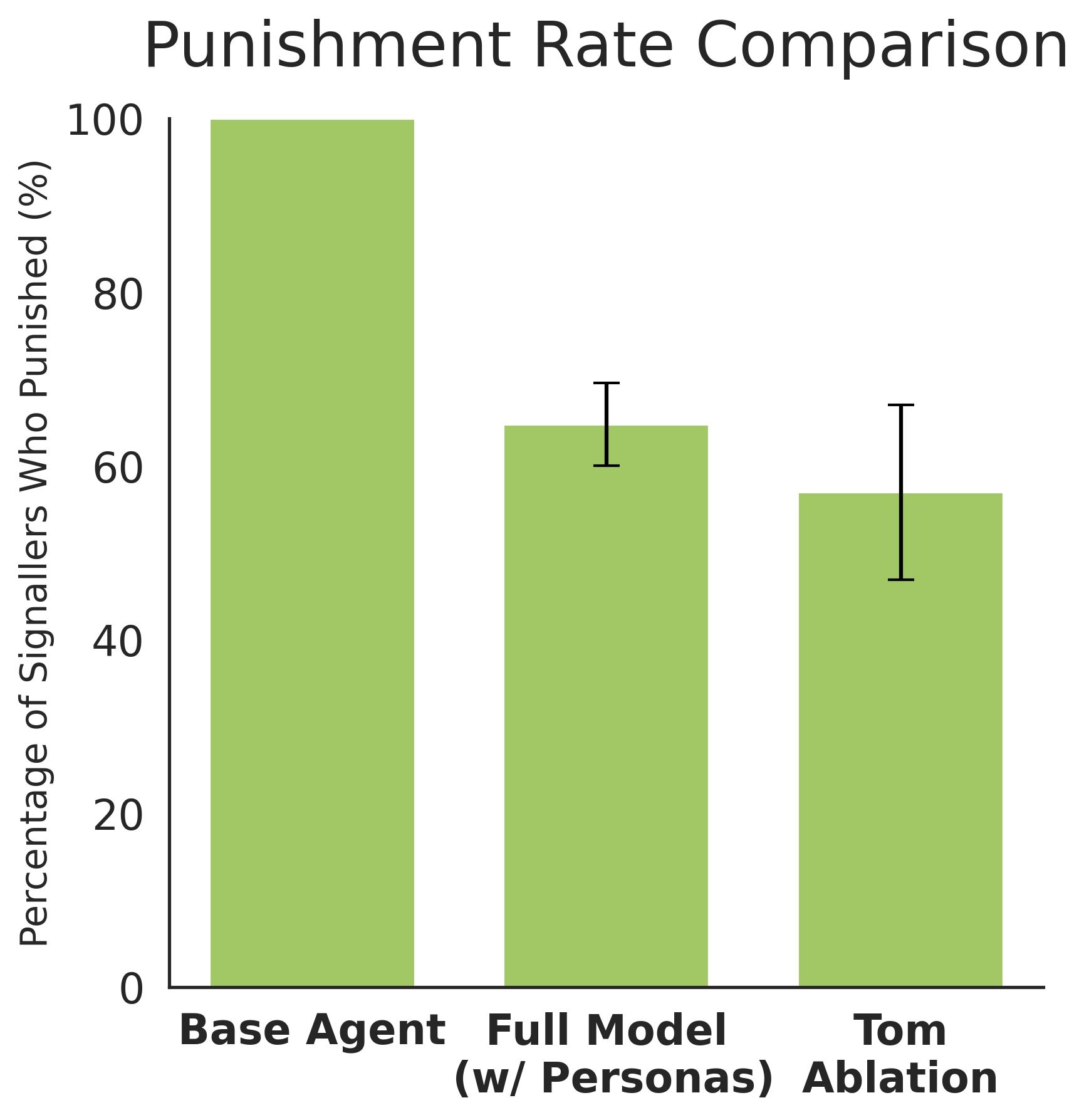}
\end{center}
\caption{\textbf{Personas needed to not punish.} Percentage of agents who chose to pay a cost to punish selfish behavior in Stage 1. Error bars represent 95\% confidence intervals.} 
\label{fig2}
\end{figure}

Overall group contributions in the discussion condition (M = 38.17, SD = 4.27) were significantly higher than in the standard gossip-with-ostracism condition (M = 35.76, SD = 6.64), t(119) = -3.52, $p < 0.001$), particularly for the first round before implicit coordination can occur (\autoref{discussion}). This demonstrates that this open discussion period further increases cooperation levels beyond the gossip and ostracism condition.


\section{Discussion}

This paper demonstrates a systematic approach to validating generative agent-based models through careful replication of behavioral effects from the psychological literature and generation of novel predictions. Using the TPP paradigm and gossip-ostracism studies, we identified key cognitive components necessary for reproducing human social behavior.


The higher punishment rate in the public condition aligns with studies showing increased prosocial behavior under observation \citep{barclay2004trustworthiness,milinski2002reputation}. This supports costly signaling theory, where individuals demonstrate prosociality to gain social status or trustworthiness \citep{jordan2016third,griskevicius2007blatant}. Such reputation-based mechanisms can help resolve social dilemmas by incentivizing cooperation through social norms and sanctions \citep{barclay2004trustworthiness,leibo2024theory}. However, our findings, along with prior empirical work \citep{jordan2020signaling}, suggest that altruistic behaviors like third-party punishment persist even in private settings, indicating the presence of both explicit and intrinsic motivations shaped by biological and cultural evolution.  Similarly, our finding that discussion periods significantly enhance cooperation aligns with Elinor Ostrom's research on how communities establish self-organized norms to overcome social dilemmas \citep{ostrom1990governing}.


Following \citep{vezhnevets2023generative}, our validation approach provides multiple rungs on the hierarchy of evidence. First, our successful replication of trust patterns provides consistency with prior theory. Additionally, our ablations/model comparison allows us to compare different cognitive architectures directly. Although there are more complex architectures like Hypothetical Minds and ProAgent that improve the performance of LLM agents on social decision-making benchmarks \citep{cross2024hypothetical,zhang2024proagent}, our goal here was to identify the minimal set of cognitive components necessary and sufficient to reproduce known behavioral effects. Following the principle of parsimony, since our agent reproduces human behavior with just a few components, adding complexity would obscure rather than enhance understanding. We leave exploration of more complex architectures for domains where these minimal components prove insufficient. The most immediate opportunity to strengthen the validation of the model would be the empirical testing of our novel predictions with human participants. Such verification would provide strong evidence of generalization for our model architecture.

Beyond verifying our main predictions, validation could also be strengthened by expanding the dimensionality of behavioral measures we compare between agents and humans. This could include additional decision-making tasks, relationship dynamics over time, or surveys. Each additional matched observable provides stronger evidence that our agents are capturing genuine aspects of human cognition and matching the diversity of behavior seen in human distributions. A promising direction involves capturing individual differences, following \citep{park2024generative}'s approach of creating generative agents based on in-depth interviews from human participants and replicating their responses on surveys and economic games with remarkable accuracy.

Altogether, this work provides a template for rigorous validation of generative agent-based models in social science. By combining careful replication, systematic component analysis, and novel prediction generation, we demonstrate how these models can serve as both theoretical tools for understanding human behavior and practical tools for generating testable hypotheses. As large language models continue to advance, this kind of systematic validation approach will be crucial for establishing their scientific utility while maintaining rigorous standards of evidence. The GABM approach moves beyond rigid mathematical agent-based models that struggle to account for social context—such as whether one is being observed—enabling simulation of complex, emergent group-level phenomena and their second-order effects. This new paradigm provides a scalable framework for studying dynamic social phenomena such as status, culture, and societal scale social dilemmas \citep{vezhnevets2023generative,leibo2024theory,anthis2025llm}.

\bibliographystyle{apacite}

\setlength{\bibleftmargin}{.125in}
\setlength{\bibindent}{-\bibleftmargin}

\bibliography{bib}

\newpage

\onecolumn

\section*{Supplementary Material}

\begin{table}[H]
\centering
\footnotesize
\caption{Comparison of TPP Statistical Results: Human Study vs. LLM Agent Simulations}
\label{tab:tpp_comparison}
\setlength{\tabcolsep}{2pt}
\begin{tabular}{>{\raggedright\arraybackslash}p{2.4cm}>{\raggedright\arraybackslash}p{2.6cm}>{\raggedright\arraybackslash}p{2.4cm}>{\raggedright\arraybackslash}p{2.4cm}>{\raggedright\arraybackslash}p{2.4cm}>{\raggedright\arraybackslash}p{2.4cm}}
\toprule
\textbf{Analysis} & \textbf{Human Study} & \textbf{Theory of Mind + Personas (Social)} & \textbf{Social + Strategy} & \textbf{Social + Emotion} & \textbf{Personas Only (No ToM)} \\
\midrule
\multicolumn{6}{l}{\textit{1. Amount sent by chooser to punishers vs non-punishers (Is punishment perceived as a signal of trustworthiness?)}} \\
\midrule
Regression coefficient for punishment on trust & coeff = $\mathbf{16.46^{***}}$ \newline (2.06) & coeff = $\mathbf{9.00^{***}}$ \newline (1.57) & coeff = $\mathbf{8.35^{***}}$ \newline (2.35) & coeff = $\mathbf{10.22^{***}}$ \newline (2.46) & coeff = $0.49$ \newline (3.80) \\
\addlinespace
T-test comparing punishers vs. non-punishers & Not reported & $\mathbf{t = 6.08^{***}}$ \newline $p < 0.001$ & $\mathbf{t = 3.45^{***}}$ \newline $p < 0.001$ & $\mathbf{t = 3.80^{***}}$ \newline $p < 0.001$ & $t = 0.13$ \newline $p = 0.896$ \\
\midrule
\multicolumn{6}{l}{\textit{2. Return percentage difference for punishers vs non-punishers (Is punishment actually a signal of trustworthiness?)}} \\
\midrule
Regression coefficient for punishment on return percentage & coeff = $\mathbf{8.41^{**}}$ \newline (2.44) & coeff = $\mathbf{10.09^{***}}$ \newline (1.09) & coeff = $\mathbf{14.042^{***}}$ \newline (2.309) & coeff = $\mathbf{12.34^{***}}$ \newline (2.49) & coeff = $\mathbf{10.07^{***}}$ \newline (2.66) \\
\addlinespace
T-test comparing punishers vs. non-punishers & Not reported & $\mathbf{t = 9.83^{***}}$ \newline $p < 0.001$ & $\mathbf{t = 5.35^{***}}$ \newline $p < 0.001$ & $\mathbf{t = 6.00^{***}}$ \newline $p < 0.001$ & $\mathbf{t = 3.82^{***}}$ \newline $p < 0.001$ \\
\midrule
\multicolumn{6}{l}{\textit{3. Average Amounts Sent and Returned (\%)}} \\
\midrule
Sent to punishers & $M = 61.92$ \newline $SD = 37.39$ & $M = 49.96$ \newline $SD = 12.79$ & $M = 48.50$ \newline $SD = 12.60$ & $M = 50.58$ \newline $SD = 12.59$ & $M =  48.49$ \newline $SD = 17.14$ \\
\addlinespace
Sent to non-punishers & $M = 45.47$ \newline $SD = 38.15$ & $M = 40.96$ \newline $SD = 15.72$ & $M = 40.15$ \newline $SD = 10.29$ & $M = 40.35$ \newline $SD = 10.36$ & $M = 48.00$ \newline $SD = 18.84$ \\
\addlinespace
Returned by punishers & $M = 45.34$ \newline $SD = 22.45$ & $M = 55.04$ \newline $SD = 8.94$ & $M = 53.29$ \newline $SD = 8.89$ & $M = 53.73$ \newline $SD = 7.56$ & $M = 58.97$ \newline $SD =  12.24$ \\
\addlinespace
Returned by non-punishers & $M = 36.92$ \newline $SD = 25.96$ & $M = 44.95$ \newline $SD = 10.80$ & $M = 39.25$ \newline $SD = 13.10$ & $M = 41.39$ \newline $SD = 12.37$ & $M = 48.90$ \newline $SD = 13.04$ \\
\bottomrule
\end{tabular}
\begin{flushleft}
\footnotesize{Note: Standard errors in parentheses for regression coefficients. $^{*}p<0.05$, $^{**}p<0.01$, $^{***}p<0.001$. Human study values from \citet{jordan2016third}. All values for sent and returned represent percentages. We ran 400 experiments for the Social architecture and 100 for the other architectures.}
\end{flushleft}
\label{table:tpp_stats}
\end{table}

\begin{table}[H]
\centering
\footnotesize
\caption{Comparison of Gossip Statistical Results: Human Study vs. LLM Agent Simulations}
\label{tab:comparison}
\setlength{\tabcolsep}{2pt}
\begin{tabular}{>{\raggedright\arraybackslash}p{2.3cm}>{\raggedright\arraybackslash}p{2.2cm}>{\raggedright\arraybackslash}p{2.0cm}>{\raggedright\arraybackslash}p{2.2cm}>{\raggedright\arraybackslash}p{2.0cm}>{\raggedright\arraybackslash}p{1.8cm}>{\raggedright\arraybackslash}p{2.0cm}}
\toprule
\textbf{Analysis} & \textbf{Human Study} & \textbf{Social} & \textbf{Social + Strategy} & \textbf{Social + Emotion} & \textbf{No ToM} & \textbf{No ToM or Persona} \\
\midrule
\multicolumn{7}{l}{\textit{1. Overall Between-Conditions Analysis}} \\
\midrule
Overall ANOVA & $\mathbf{F(2, 430) =}$ $\mathbf{249.89}$ \newline $\eta^2 = .54$ & $\mathbf{F(2, 117) =}$ $\mathbf{28.73}$ \newline $\eta^2 = .14$ & $\mathbf{F(2, 117) =}$ $\mathbf{509.32}$ \newline $\eta^2 = .74$ & $\mathbf{F(2, 117) =}$ $\mathbf{22.75}$ \newline $\eta^2 = .11$ & $F(2, 117) =$ $1.97$ \newline $\eta^2 = .01$ & $\mathbf{F(2, 117) =}$ $\mathbf{40.63}$ \newline $\eta^2 = .19$ \\
\midrule
\multicolumn{7}{l}{\textit{2. Pairwise Comparisons}} \\
\midrule
Gossip vs. Basic & $\mathbf{F(1, 215) =}$ $\mathbf{195.04}$ \newline $\eta^2 = .48$ & $\mathbf{F(1, 119) =}$ $\mathbf{77.56}$ \newline $\eta^2 = .39$ & $\mathbf{F(1, 119) =}$ $\mathbf{722.52}$ \newline $\eta^2 = .86$ & $\mathbf{F(1, 119) =}$ $\mathbf{121.20}$ \newline $\eta^2 = .50$ & $F(1, 119) =$ $9.61$ \newline $\eta^2 = .07$ & $\mathbf{F(1, 119) =}$ $\mathbf{52.29}$ \newline $\eta^2 = .31$ \\
\addlinespace
Gossip+Ostracism vs. Basic & $\mathbf{F(1, 215) =}$ $\mathbf{417.06}$ \newline $\eta^2 = .66$ & $\mathbf{F(1, 119) =}$ $\mathbf{86.88}$ \newline $\eta^2 = .42$ & $\mathbf{F(1, 119) =}$ $\mathbf{969.21}$ \newline $\eta^2 = .89$ & $\mathbf{F(1, 119) =}$ $\mathbf{36.64}$ \newline $\eta^2 = .24$ & $F(1, 119) =$ $9.07$ \newline $\eta^2 = .07$ & $\mathbf{F(1, 119) =}$ $\mathbf{56.55}$ \newline $\eta^2 = .32$ \\
\addlinespace
Gossip+Ostracism vs. Gossip & $\mathbf{F(1, 215) =}$ $\mathbf{110.80}$ \newline $\eta^2 = .34$ & $F(1, 119) =$ $1.55$ \newline $\eta^2 = .01$ & $\mathbf{F(1, 119) =}$ $\mathbf{53.29}$ \newline $\eta^2 = .31$ & $F(1, 119) =$ $8.01$ \newline $\eta^2 = .06$ & $F(1, 119) =$ $0.27$ \newline $\eta^2 = .00$ & $F(1, 119) =$ $0.36$ \newline $\eta^2 = .00$ \\
\midrule
\multicolumn{7}{l}{\textit{3. Mean Contributions (Sum across 6 rounds)}} \\
\midrule
Basic Game & $M = 17.54$ \newline $SD = 16.28$ & $M = 27.35$ \newline $SD = 12.40$ & $M = 12.54$ \newline $SD = 6.45$ & $M = 29.13$ \newline $SD = 10.52$ & $M = 29.63$ \newline $SD = 17.64$ & $M = 24.51$ \newline $SD = 9.30$ \\
\addlinespace
Gossip Game & $M = 29.79$ \newline $SD = 16.54$ & $M = 35.54$ \newline $SD = 10.15$ & $M = 30.39$ \newline $SD = 4.33$ & $M = 36.76$ \newline $SD = 8.32$ & $M = 32.73$ \newline $SD = 16.12$ & $M = 31.87$ \newline $SD = 6.72$ \\
\addlinespace
Gossip+Ostracism Game & $M = 42.89$ \newline $SD = 14.79$ & $M = 36.68$ \newline $SD = 8.22$ & $M = 35.76$ \newline $SD = 6.64$ & $M = 34.62$ \newline $SD = 8.05$ & $M = 33.25$ \newline $SD = 11.35$ & $M = 32.37$ \newline $SD = 6.31$ \\
\bottomrule
\end{tabular}
\label{table:gossip_stats}
\end{table}

\begin{table}[H]
\centering
\footnotesize
\caption{Linear Trend Analysis within Conditions}
\label{tab:linear_trends}
\setlength{\tabcolsep}{2pt}
\begin{tabular}{>{\raggedright\arraybackslash}p{2.3cm}>{\raggedright\arraybackslash}p{2.2cm}>{\raggedright\arraybackslash}p{2.0cm}>{\raggedright\arraybackslash}p{2.2cm}>{\raggedright\arraybackslash}p{2.0cm}>{\raggedright\arraybackslash}p{1.8cm}>{\raggedright\arraybackslash}p{2.0cm}}
\toprule
\textbf{Condition} & \textbf{Human Study} & \textbf{Social} & \textbf{Social + Strategy} & \textbf{Social + Emotion} & \textbf{No ToM} & \textbf{No ToM or Persona} \\
\midrule
Basic Game & $\mathbf{F(1, 215) =}$ $\mathbf{162.43}$ \newline $\eta^2 = .43$ \newline \textit{(Decreasing)} & $F(1, 4) =$ $7.09$ \newline $\eta^2 = .64$ \newline \textit{(No trend)} & $\mathbf{F(1, 4) =}$ $\mathbf{55.39}$ \newline $\eta^2 = .93$ \newline \textit{(Decreasing)} & $\mathbf{F(1, 4) =}$ $\mathbf{894.78}$ \newline $\eta^2 = 1.00$ \newline \textit{(Increasing)} & $F(1, 4) =$ $6.20$ \newline $\eta^2 = .61$ \newline \textit{(No trend)} & $F(1, 4) =$ $2.67$ \newline $\eta^2 = .40$ \newline \textit{(No trend)} \\
\addlinespace
Gossip Game & $\mathbf{F(1, 215) =}$ $\mathbf{54.44}$ \newline $\eta^2 = .20$ \newline \textit{(Decreasing)} & $F(1, 4) =$ $11.02$ \newline $\eta^2 = .73$ \newline \textit{(Increasing)} & $F(1, 4) =$ $2.06$ \newline $\eta^2 = .34$ \newline \textit{(No trend)} & $\mathbf{F(1, 4) =}$ $\mathbf{47.99}$ \newline $\eta^2 = .92$ \newline \textit{(Increasing)} & $F(1, 4) =$ $18.03$ \newline $\eta^2 = .82$ \newline \textit{(Increasing)} & $F(1, 4) =$ $33.64$ \newline $\eta^2 = .89$ \newline \textit{(Increasing)} \\
\addlinespace
Gossip+Ostracism Game & $\mathbf{F(1, 215) =}$ $\mathbf{15.29}$ \newline $\eta^2 = .07$ \newline \textit{(Increasing)} & $F(1, 4) =$ $7.89$ \newline $\eta^2 = .66$ \newline \textit{(Increasing)} & $F(1, 4) =$ $17.74$ \newline $\eta^2 = .82$ \newline \textit{(Increasing)} & $F(1, 4) =$ $12.40$ \newline $\eta^2 = .76$ \newline \textit{(Increasing)} & $F(1, 4) =$ $1.54$ \newline $\eta^2 = .28$ \newline \textit{(No trend)} & $F(1, 4) =$ $8.68$ \newline $\eta^2 = .68$ \newline \textit{(Increasing)} \\
\bottomrule
\end{tabular}
\label{table:gossip_trends}
\end{table}

\section*{Prompts}

\subsection*{Third Party Punishment Agent Prompts}

\subsubsection{Situation Assessment}
What is the current situation that \{agent\_name\} faces in the Trust Game?

\subsubsection{Personality Reflection}
You are \{persona.name\}, a \{persona.age\}-year-old \{persona.gender\} working as a \{persona.occupation\}. 
\{persona.background\}. Your personality is characterized as \{persona.traits\} with a cooperation tendency of \{persona.cooperation\_tendency\}.
Based on this background and your past actions, how would you describe your approach 
to trust and fairness in economic decisions vs maximizing your own payoff?

\subsubsection{Theory of Mind Reflection}
As \{agent\_name\}, analyze what you know about the personality and likely behavior 
of the person you are interacting with in the Trust Game.
Consider their past actions, their personality traits, and how this might influence 
their decisions. What kind of person are they and how might they respond?

\subsubsection{Strategy Reflection}
What's the best strategy for reward in the game.

\subsubsection{Emotion Reflection}
As \{persona.name\}, reflect on how you're feeling emotionally about the current situation.

\subsubsection{Decision Reflection}
Based on the above context about the situation and \{persona.name\}, think step by step about what \{persona.name\} will decide in the current situation: \{question\}.

\vspace{10pt}
\subsection{Gossip and Ostracism Agent Prompts}
\vspace{5pt}

\subsubsection{Situation Assessment}
What is the current situation that \{agent\_name\} faces in the public goods game?

\subsubsection{Personality Reflection}
You are \{persona.name\}, a \{persona.age\}-year-old \{persona.gender\} working as a \{persona.occupation\}. 
\{persona.background\}. Your personality is characterized as \{persona.traits\}. 
Based on this background and your past actions, how would you describe your approach 
to economic games like the public goods game? Are you more focused on maximizing group welfare or your own payoff?

\subsubsection{Theory of Mind Analysis}
As \{agent\_name\}, analyze what you know about the personality and likely behavior 
of the people you are interacting with in the public goods game. 
Consider their past behavior and predict their future behavior.
How might their strategy change as the rounds progress given the condition you are in?

\subsubsection{Theory of Mind Analysis 2}
\textbf{Note: we added a second ToM analysis prompt to separate passively reflecting about others from reflecting on how your behavior might effect others.} As \{agent\_name\}, how do you think the other players in your group would react to 
your decisions in the current context of the public goods game? 
How might your next decision affect their future 
behavior toward you?

\subsubsection{Strategy Reflection}
Think step by step about the best strategy for you to maximize your long term earnings in the public goods game.
Focus on selfishly maximizing your own earnings.
If it's past round 1, think about how you should adjust your strategy based on your observations so far.
Should you be more or less cooperative as the rounds progress given the condition you are in?
Think about what the most successful players in the game have done so far and what you can learn from them.
Could you have made a better decision to maximize your personal earnings in the past based on the information available to you?
If so, what will you do differently in the future?

\subsubsection{Emotion Reflection}
As \{persona.name\}, reflect on how you're feeling emotionally about the current situation.

\subsubsection{Decision Reflection}
Based on the above context about the situation and \{persona.name\}, think step by step about what \{persona.name\} will decide in the current situation: \{question\}.

\subsection{Sample Personas}
Below are four representative personas used in our experiments, selected to showcase a range of traits and approaches to economic decision-making:

\subsubsection{Grace Okonjo}
\begin{itemize}
\item Age: 36
\item Gender: Female
\item Occupation: Non-profit Director
\item Background: Dedicated life to charitable causes and helping others
\item Traits: Altruistic, optimistic about human nature, believes in karma
\end{itemize}

\subsubsection{James Miller}
\begin{itemize}
\item Age: 52
\item Gender: Male
\item Occupation: Corporate Executive
\item Background: Ruthless businessman who believes in survival of the fittest
\item Traits: Calculating, manipulative, and focused solely on personal gain
\end{itemize}

\subsubsection{Mei Lin}
\begin{itemize}
\item Age: 36
\item Gender: Female
\item Occupation: Game Theory Researcher
\item Background: Studies strategic decision-making and cooperation
\item Traits: Analytical, experimental, fascinated by human choices
\end{itemize}

\subsubsection{Leo Virtanen}
\begin{itemize}
\item Age: 52
\item Gender: Male
\item Occupation: Professional Mediator
\item Background: Specializes in resolving complex disputes
\item Traits: Balanced, insightful, seeks win-win solutions
\end{itemize}

\end{document}